\documentclass[12pt]{article}

\usepackage{amsmath,amssymb,subfigure,amsthm,mathtools,enumerate}\usepackage[mathscr]{eucal}
\usepackage{graphicx}
\usepackage{color} 
\newcommand{\R}{\mathbb{R}}

\newcommand{\Z}{\mathbb{Z}}
\newcommand{\N}{\mathbb{N}}
\newcommand{\E}{\mathbb{E}}
\renewcommand{\P}{\mathbb{P}}

\newtheorem{theorem}{Theorem}[section]

\newtheorem{proposition}[theorem]{Proposition}

\newtheorem{corollary}[theorem]{Corollary}

\title{Lower bounds for the error decay incurred by coarse quantization schemes}
\author{Felix Krahmer \footnotemark[1] \and Rachel Ward  \footnotemark[2]}

\begin{document}
\renewcommand{\thefootnote}{\fnsymbol{footnote}}
 \footnotetext[1]{
   Hausdorff Center for Mathematics, Universit{\"a}t Bonn, Bonn, Germany
   }
\footnotetext[2]{
    Courant Institute of Mathematical Science, New York University, New York, NY, USA
    }   
\renewcommand{\thefootnote}{\arabic{footnote}}

\maketitle

\begin{abstract} 
Several analog-to-digital conversion  methods for bandlimited signals used in applications, such as $\Sigma\Delta$ quantization schemes,
employ coarse quantization coupled with oversampling.  The standard mathematical model for the error accrued from such methods measures the performance of a given scheme by the rate at which the associated reconstruction error decays as a function of the oversampling ratio $\lambda$. It was recently shown that exponential accuracy of the form $O(2^{-\alpha \lambda})$ can be achieved by appropriate one-bit Sigma-Delta modulation schemes. However, the best known achievable rate constants $\alpha$ in this setting differ significantly from the general information theoretic lower bound. In this paper, we provide the first lower bound specific to coarse quantization, thus narrowing the gap between existing upper and lower bounds.  In particular, our results imply a quantitative correspondence between the maximal signal amplitude and the best possible error decay rate. Our method draws from the theory of large deviations.
\end{abstract}

\section{Introduction}
Many signals of practical engineering interest are naturally produced in analog form; at the same time, it is becoming more efficient and robust to store and transmit signals in digital form.  Therefore, the study of accurate and tractable methods for analog-to-digital (A/D) conversion, or the approximation of real-valued signals using a finite alphabet,  is of great importance in modern signal processing.  

In the setting of A/D conversion, the signal of interest $x(t)$ is often modeled as a bounded bandlimited function.  According to the well-known Shannon-Nyquist sampling theorem, such functions are completely determined by their values $x_n= x(\frac{n}{\lambda})$ sampled at frequency $\lambda$ greater than the signal bandwidth. The original signal can be reconstructed from these samples using convolutional decoding of the form $x(t) =\frac{1}{\lambda} \sum_{n \in \mathbb{Z}} x_n g(t - \frac{n}{\lambda}).$ For exact equality,  the Fourier transform of the function $g \in L^{\infty}(\mathbb{R})$ needs to approximate the characteristic function of the frequency support of $x(t)$; in particular, $\widehat g$ needs to have compact support.  In that case, the reconstruction formula represents an ideal low-pass filter.  Conversion between analog and digital representations for $x(t)$ may be achieved by replacing the input sequence $( x_n)$ by a sequence $(q_n)$ of \emph{quantized values} chosen from a finite set such that the signal 
\begin{equation}
\label{continuous}
\widetilde{x}(t) = \sum_{n \in {\Z}} q_n g\big( t - \frac{n}{\lambda} \big)
\end{equation}
 formed by replacing the $x_n$'s with the $q_n$'s yields a good approximation of $x$.  In applications, one is often forced to approximate the ideal low-pass filter $g$ by a filter $\varphi$ satisfying additional constraints, as for example compact (time) support.  
 
In addition, one sometimes restricts attention to recovering the values $x_j$ on the sampling grid only.  Consequently, such a quantization scheme fixes a finite-length reconstruction filter $\varphi_n$, and approximate recovery is then obtained if
\begin{equation}
\label{discrete}
x_j \approx \tilde{x}_j = \sum \varphi_{j-n} q_n.
\end{equation} 
In this paper we will focus on the continuous scenario \eqref{continuous}, but we will allow for (almost) arbitrary reconstruction kernels $\varphi$.
Similar techniques extend to corresponding results for the discrete scenario \eqref{discrete}.  
 
\subsection*{Quantization schemes employed in practice}  In \emph{pulse code modulation}, the sampling frequency $\lambda$ is close to the critical sampling frequency, and the quantized value $q_n$ is taken to be a truncated binary representation of the sample $x_n$. To increase the accuracy of this approximation, one takes longer binary expansions of each sample.  In particular, if $m$ bits are allotted to each truncated binary expansion, then the distortion $\| x - \widetilde{x} \|_{L^\infty}$ decreases like $O(2^{-m})$. 

On the other hand, the set of admissible values for $q_n$ in \emph{oversampled coarse quantization} methods is restricted to a fixed alphabet ${\cal A}$ of reasonably small size, and  more accurate approximations are obtained by increasing the sampling rate $\lambda$.  In the extreme case of one-bit quantization, one chooses the alphabet ${\cal A}_1 = \{ -1, +1 \}$.  For $K$-bit quantization, the $q_n$ are taken from the set ${\cal A}_K$ consisting of $2^K$ evenly spaced values in the closed interval $[-1,1]$.  The number of bits spent per unit  time interval in this setting is $m = \lambda \log_2{| \cal{A}_K |} = \lambda K$.   From the viewpoint of circuit engineering, oversampled coarse quantization is associated to low-cost analog hardware, because increasing the sampling rate is cheaper than refining the quantization.  Consequently, oversampling data converters are often used for low to medium-bandwidth signals, such as audio signals \cite{SD} and, more recently, for wireless communication \cite{wireless}.  Further advantages of oversampled coarse quantization methods include a built-in redundancy and robustness against errors resulting from imperfections in the analog circuit implementation.  This robustness comes as a consequence of the more `democratic' distribution of bit significance in the reconstruction formula, see \cite{DC02}; in the extreme case of one-bit quantization, the individual bits $q_n \in \{-1,1\}$ carry \emph{equal} significance.

\subsection*{Our work in relation to prior advances} In this paper, we show that these advantages of coarse quantization come with the price of sub-optimal accuracy of the resulting convolutional approximation.  It is well-known (see, for example, \cite{KT}, \cite{S}) that no quantization scheme spending $m$ bits per Nyquist interval can beat the error decay of $O(2^{-m})$ achieved by pulse code modulation.  This optimal rate of decay is not possible for coarse quantization in the discrete setting \eqref{discrete}, following the work of Calderbank and Daubechies \cite{DC02}.  Until now, tighter lower bounds for coarse quantization are only available under the white noise hypothesis, where one assumes that the quantization error $x_n - q_n$ is distributed like Gaussian white noise, and in conjunction with additional technical assumptions \cite{SQG}.   In contrast, the lower bounds we shall provide hold for \emph{any} $K$-bit quantization scheme, without any additional assumptions, and independent of the encoding algorithm used to generate the $q_n$.

As the main contribution of this paper, we provide an explicit lower bound on the error decay achievable by $K$-bit quantization.  Normalizing such that the $q_n$'s are chosen from an evenly spaced alphabet with endpoints $-1$ and $1$, and such that the bandlimited functions of interest are bounded in amplitude by $\mu < 1$, we will show that the rate of decay of $\| \widetilde{x}_{\lambda} - x \|_{\infty} $ is bounded below by $O\big( 2^{-\alpha m} \big)$, where $\alpha = 1 - K^{-1} (1 - h(\frac{1+\mu}{2}))$, and $h$ is the \emph{unbiased binary entropy function } $h(u) = (1 - u)\log_2{(1 - u)} + u\log_2{u}.$  In fact, the best known upper bounds for $K$-bit quantzation are also of the form $O(2^{-r m})$.  Such a bound was first achieved via a construction by G\"unt\"urk \cite{S} of a a family of one-bit $\Sigma \Delta$ quantization schemes.  These constructions were later refined by Deift, G\"unt\"urk, Krahmer \cite{DGK10}, yielding the rate constant $r \approx 0.102$.  As this rate constant is achieved only over input signals of maximal amplitude $\mu \leq .05$, this upper bound does not stand in contradiction to our lower bound, which implies in particular that the best possible rate constant tends to zero as  $\mu \rightarrow 1$.  

\subsection*{Organization of the paper} After precisely setting up the problem and clarifying our notation in Section 2, we summarize our results in Section 3. In Section 4, we recall important concepts and results from the theory of large deviations.  In that section, we also recall results from the theory of Banach spaces which we use in the proof of our main theorem, which is presented in Section 5.


\section{Notation and setup}
Before continuing, let us survey the notation used in this paper.  We use the Landau O-notation $f(x) = O(h(x))$ (and $f(x) = o(h(x))$)  to imply that for some $M>0$ (or any $M>0$, respectively), there exists a real number $u_0$ such that  $|f(u)| \leq M |h(u)|$  for all $u \geq {u}_0$.  Let ${\cal S}$ denote the Schwartz space of rapidly decreasing functions on $\mathbb{R}$.  For the Fourier transform, we use the normalization
\begin{equation}
\widehat{x}(\omega) := \int_{-\infty}^{\infty} x(t) \exp{(-2\pi i \omega t)} dt.
\label{fourier}
\end{equation}
We define the class ${\cal B}_{\Omega}(\R)$ of $\Omega$-bandlimited functions to be the space of real-valued continuous functions in $L^{\infty}(\mathbb{R})$ whose Fourier transforms (in the distributional sense) have support contained in $[-\Omega/2 , \Omega/2]$.  Henceforth, we will normalize $\Omega = 1$. 
The classical sampling theorem for bandlimited functions states that if $\lambda > 1$, then any function $x$ in the class ${\cal B}_{1}(\R)$ and having bounded amplitude can be recovered from its samples $\{ x (\frac{n}{\lambda}) \}_{n \in \mathbb{Z}}$ as a weighted sum of translates of an averaging kernel $g \in L^1(\mathbb{R})$ via the formula
\begin{equation}
x(t) = \frac{1}{\lambda} \sum_{n \in \mathbb{Z}} x \Big( \frac{n}{\lambda} \Big) g \Big( t - \frac{n}{\lambda} \Big), 
\label{reconstruct}
\end{equation}  
where $g$ is any kernel whose Fourier transform satisfies
\begin{eqnarray}
\label{varphi}
\widehat{g}(\omega) &=&  \left\{ \begin{array}{ll}1, & \textrm{if } | \omega | \leq \pi  \\
0, & \textrm{if } | \omega | \geq \lambda_0\pi 
\end{array} \right.
\end{eqnarray}
for some arbitrary $\lambda_0$ with $\lambda \geq \lambda_0 > 1$.  Note that with such a $g$, the reconstruction formula \eqref{reconstruct} describes an ideal low-pass filter.   Note also that any such kernel $g$ with finite frequency support must have infinite (time) support, according to the uncertainty principle.   Such ideal filters with infinite-support are cumbersome to construct, and in practice are often approximated by kernels having finite support.   In this case, the reconstruction formula \eqref{reconstruct} holds at most approximately.  A priori it is not clear that this approximation always has a negative effect on the accuracy of the associated quantization schemes.  For this reason, in the subsequent analysis we will not restrict the choice of the filter by more than a simple smoothness condition.  We will use, however, the normalization arising naturally in the ideal case. There one has by \eqref{varphi} that
\begin{equation}
\label{gnorm}
\int g(t) dt = \widehat{g}(0) = 1;
\end{equation}
we adapt this normalization for general kernels $\varphi$.  
\\
\\
A {\em $K$-bit quantization scheme} assigns, to each input function $x$ and to each sampling rate $\lambda \geq \lambda_0$, a sequence of evenly-spaced $q_n^{\lambda}$ from an alphabet ${\cal A}_K$ of size $| {\cal A}_K | = 2^K$ in such a way that the approximation
\begin{equation}\label{eq:approx}
\widetilde{x}_{\lambda}(t) = \frac{1}{\lambda}\sum_{n \in \mathbb{Z}} q_n^{\lambda} \varphi\Big(t - \frac{n}{\lambda}\Big)
\end{equation}
approaches $x(t)$ as $\lambda \rightarrow \infty$. 
Consequently, the approximation quality resulting from a particular sequence $\{q^\lambda_n\}_{n\in\Z}$ of quantized values together with a reconstruction kernel $\varphi$ is commonly assessed by the reconstruction error,
\begin{equation}\label{eq:norm}
e^K_{\lambda, q^\lambda}(t) := x(t) - \frac{1}{\lambda} \sum_{n \in \mathbb{Z}} q_n^{\lambda} \varphi(t - \frac{n}{\lambda}),  \hspace{10mm} q_n^{\lambda} \in {\cal A}_K
\end{equation}
and its supremum norm. We shall normalize the $K$-bit quantization alphabet ${\cal A}_{K}$ so as to have extreme values $+1$ and $-1$.  

With this normalization on the alphabet in place and the kernel normalization \eqref{gnorm}, the approximate reconstruction in \eqref{eq:approx} is essentially a weighted local average of the $q_n^{\lambda}$. Hence we can not expect good approximation for $\| x \|_{L^{\infty}} > 1$.   For this reason, we fix $\mu< 1$ and work with the space $\mathcal{B}^\mu_1(\mathbb{R}, \mu)$ defined to be the class of functions in $\mathcal{B}_1(\mathbb{R})$ 
with amplitude bounded by $\mu$ on the whole real line.  
Thus we will study
\begin{equation}
\label{eopt}
E^{\mu}_{K}(\lambda) := \sup_{x \in {\cal B}_1(\mathbb{R},\mu)} \inf_{q_n^{\lambda} \in {\cal A}_K} \| e^K_{\lambda} \|_{L^\infty}.
\end{equation}



\section{Summary of results}


Our main result concerns a lower bound on the rate of decay for $K$-bit quantization of bandlimited functions in terms of the maximal amplitude $\mu$:
\begin{theorem}
\label{thm:main}
Consider a $K$-bit quantization scheme associated to a reconstruction kernel $\varphi \in {\cal S}$, normalized so that $\int \varphi(t) dt = 1$.  If the optimal rate of decay for such a scheme satisfies $E^{\mu}_{K}(\lambda)=O(2^{-\alpha K \lambda})$, then 
$$
\alpha \leq 1- K^{-1} \left( 1- h\left(\frac{1+\mu}{2}\right)\right) ,$$
where $h(p) = p \log_2{p} + (1-p) \log_2{(1-p)}$ is the binary entropy function.
\end{theorem}

Theorem \ref{thm:main} represents a quantitative improvement over the general lower bound, which for $K$-bit quantization reads
\begin{equation}
\label{eqn:prev}
E^\mu_{K} \geq O(2^{-K\lambda}),
\end{equation}
as well as over the corresponding strict inequality in the discrete case (as mentioned above).

\begin{figure}[h]\label{fig:ul}
\begin{center}
\includegraphics[width=8cm]{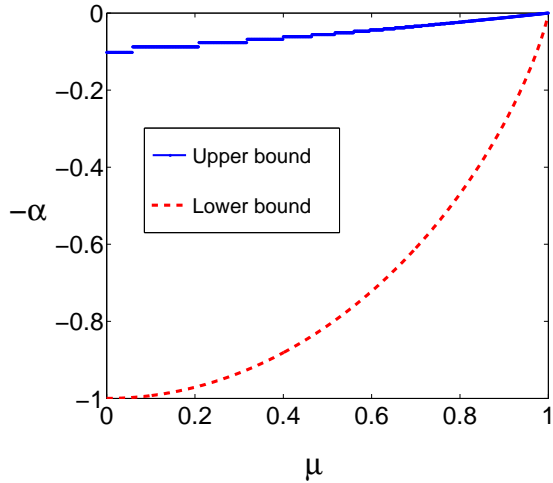}
\caption{The rate constants $\alpha$ corresponding to upper and lower bounds for the error decay as a function of $\mu$ and for $1$-bit quantization.}
\end{center}
\end{figure}

The lower bound provided in Theorem \ref{thm:main} is most markedly improved over the previous lower bound \eqref{eqn:prev} in the case of one-bit quantization, $K=1$. In this case,  the bound reduces to $\alpha \leq h(\frac{1+\mu}{2})$.  In Figure~\ref{fig:ul}, we compare our lower bound with the best-known upper bounds from \cite{DGK10} in this setting. Observe that in the limit as $\mu \rightarrow 1$, the upper and lower bounds both yield $\alpha=0$. For small $\mu$, however, there is a considerable gap between the lower bounds provided in this paper and the best-known constructive upper bounds in \cite{DGK10}. A possible explanation for that fact is that our lower bounds hold for arbitrary bit sequences, while there need not be a constructive procedure to find the optimal bit sequence from a signal. 

\subsection*{Intuition behind Theorem \ref{thm:main}} That the performance of $K$-bit quantization schemes should depend on the maximal amplitude $\mu$ can be understood as follows.  Among the $2^{KN}$ sequences of length $N$ comprised of elements $q_n \in {\cal A}_K$,  most of the sums $\sum_{n=0}^{N} q_n$ will have an average near zero.  Now the values of the reconstructed function $\widetilde{x}  = \sum_{|n| \leq N} q_n \varphi(t - \frac{n}{\lambda})$ are computed as a local average of the $q_n$'s, hence most of the possible $\widetilde{x}$ are localized near zero as well.  The larger $\mu$, the larger the function values to be represented; the disproportion increases.

\subsection*{Positive time sampling}
We note that in practice, the input signal $x(t)$ is only accessible for positive time $t \geq 0$, so one needs to reconstruct it from positive-time samples $x(\frac{n}{\lambda}), n \in \N$ only.  That is, it is more realistic to consider approximations of the form
\begin{equation}
\widetilde{x}_{\lambda}^+(t) = \frac{1}{\lambda} \sum_{n \in \N} q_n^{\lambda} \varphi(t - \frac{n}{\lambda}).
\end{equation}
Then the quantity $T_0(\lambda)$ defined above can be interpreted as a `calibration time' over which the approximation need not hold.  Accordingly, one measures the reconstruction error through
\begin{equation}
\label{Eopt2}
E_{K}^{\mu, +}(\lambda) := \sup_{x \in \cal{B}_{1}(\mathbb{R}, \mu)} \sup_{t \geq T(\lambda)} \left| x(t) - \frac{1}{\lambda}\sum_{n \in N} q_n^{\lambda} \varphi(t - \frac{n}{\lambda}) \right|.
\end{equation}
This is the same scenario as considered in, \cite{S}, so the effect of using only positive-time samples can be controlled as in that work. One obtains the following corollary to Theorem \ref{thm:main}:

\begin{corollary}
\label{positive-time}
If the optimal rate of decay for $E_K^{\mu,+}(\lambda)$ satisfies $E^{\mu,+}_K(\lambda)=O(2^{-\alpha K \lambda})$, then 
\begin{equation}
\alpha \leq 1-K^{-1} \Big( 1 - h \Big( \frac{1+\mu}{2} \Big) \Big).
\end{equation}
\end{corollary}

\section{Background}

\subsection{Inequalities from the theory of large deviations}

In order to make the intuition behind Theorem \ref{thm:main} rigorous, we need some results from the theory of large deviations for Bernoulli random variables. Recall that a Bernoulli random variable $X$ with bias $p$ takes values in the set $\{0,1\}$ with $\P(X = 1) = p$.  The {\em relative entropy} between two Bernoulli distributions with associated biases $p$ and $a$ is given  by $H= H(a,p) := a \log_2 \left(\frac{a}{p}\right) + (1-a)\log_2 \left(\frac{1-a}{1-p}\right)$.  In the particular case $p = 1/2$, the relative entropy function $H(a,1/2)$ simplifies to $H(a, 1/2) = h(a)+1$ where $h(a) = a \log_2(a) + (1-a) \log_2(1-a)$ is the binary entropy function.
For a sequence of independent Bernoulli random variables $B_j$ with bias $p$, denote by $S_n:= \sum_{j=1}^n B_j$ the sequence of their partial sums.   A basic result in the theory of large deviations for Bernoulli sums reads
\begin{proposition}\label{prop:1d}
For $p<a<1$, and for $n\in \N$, one has
\begin{equation}
P_n(a) := P\left(S_n \geq n a\right) \leq 2^{-nH}.
\end{equation}
\end{proposition}
Among any sum of independent and identically distributed (i.i.d) random variables $X_j$ supported on $[0,1]$ and with expected value $\E X_j = p$, the Bernoulli sum presents the slowest exponential rate of convergence towards zero for the probabilities of large deviation: 
\begin{proposition}\label{prop:max}
Let $X_1, X_2, ..., X_n$ be independent and identically distributed random variables on $[0,1]$ with $\mu = \mathbb{E}[X_n] = p$.  Then for $p < a < 1$, and for $n \in \N$, one has
\begin{equation}
P\left(\sum_{j=1}^n X_j \geq n a\right) \leq P_n(a) \leq 2^{-nH}.
\end{equation} 
\end{proposition}

For more details on large deviations for Bernoulli sums (including a detailed discussion of Proposition~\ref{prop:1d}), we refer the reader to \cite{AG}. A more complete introduction to the theory of large deviations can be found in \cite{Ellis}. For a proof of Proposition~
\ref{prop:max}, see \cite{Ellis} and \cite{LP}.

\subsection{Kolmogorov $\varepsilon$-entropy\label{sec:epsentropy}}
We need a few concepts from the theory of Banach spaces (cf. \cite{KT}). Let $Y$ be a Banach space and $X\subset Y$ a compact subset. A set $\{f_i\}_{i\in I}$, $f_i\in Y$,  is called an \emph{$\varepsilon$-net of $X$ in $Y$} if each $x\in X$ satisfies $\|x - f_i\|_\infty \leq \varepsilon$  for some $i \in I $.  Let $N = N_\varepsilon$ be
the smallest number of functions $f_1, \dots , f_N \in Y$ forming an $\varepsilon$-net of $X$
in $Y$.
The quantity 
\begin{equation}
\label{e-entropy}
H_\varepsilon:= \log_2  N_\varepsilon
\end{equation}
is the Kolmogorov $\varepsilon$-entropy (or metric entropy) of $X$ in $Y$. 

Recall that we use the notation $\mathcal{B}_1(I, \mu)$ be refer to the class of functions $x : I \rightarrow [-\mu,\mu]$ that are restrictions (to the interval $I$ ) of functions in $\mathcal{B}_1(\mathbb{R},\mu)$. This is a
compact subspace of $C(I )$ with respect to the norm $\|\cdot \|_\infty$.
The Kolmogorov $\varepsilon$-entropy of $B_1(I,\mu)$ in
$C(I)$ is shift invariant and can thus be denoted by  $H_\varepsilon(|I|)$.
 It is known \cite{KT} that the average Kolmogorov $\varepsilon$-entropy (per unit interval)
of this space, defined by
\begin{equation}
\bar{H}_\varepsilon := \lim_{|I|\rightarrow\infty} \frac{1}{|I|}H_\varepsilon(|I|)
\end{equation}
exists and has the asymptotic behavior
\begin{equation}\label{eq:heps}
\bar{H}_\varepsilon = (1 + o(1)) \log_2\frac{\mu}{\varepsilon}\text{\ \ as $\epsilon \rightarrow 0$ .}
\end{equation}
Note that we may rewrite $\log{(\frac{\mu}{\varepsilon})} = \log{(\frac{1}{\varepsilon})}(1 + o(1))$ as $\varepsilon \rightarrow 0$, so that the asymptotic behavior of $\bar{H}_{\varepsilon}$ is independent of $\mu$.
%
\\
\\
The average Kolmogorov $\varepsilon$-entropy of the space 
\begin{equation}
\mathcal{B}^\delta_{1} (\R,\mu) := \mathcal{B}_{1} (\R,\mu) \cap\big\{ f\in L^\infty \hspace{2mm} | \hspace{2mm} \forall x: f(t)\in [\mu-\delta,\mu]\big\}=\mu-\frac{\delta}{2}+\mathcal{B}_{1} (\R,\delta/2).
\end{equation}
has the same asymptotic behavior as that of $\mathcal{B}_{1}(\R, \mu)$. To see this, we use that adding a constant does not change the $\varepsilon$-entropy.

\section{Proof of Theorem \ref{thm:main}}
We are now equipped with the necessary tools to prove our main result, Theorem \ref{thm:main}.  We proceed by contradiction; more specifically we will show that under the assumption that $E^\mu_{K} \leq C2^{-\alpha K \lambda}$ for fixed constants $\alpha  > 1 - K^{-1} (1 - h( \frac{1+\mu}{2} ))$ and $C > 0$ and  all $\lambda > 1$, one can construct $\varepsilon$-nets for spaces of the type $\mathcal{B}^\delta_{1} (I,\mu)$ that violate the asymptotic bounds for the average Kolmogorov $\varepsilon$-entropy given in Section~\ref{sec:epsentropy}.

\subsection{An $\varepsilon$-net for the whole space $\mathcal{B}_1(I, \mu)$}
Let us restrict our attention to compact intervals of the form $I=[-a,a]$. Then, closely following \cite{S}, we introduce $T_0(\lambda)$ for all $\lambda > 1$ to be the
smallest number that satisfies
\begin{equation}\label{eq:T0}
\int\limits_{T_0(\lambda)-\frac{1}{\lambda}}^{\infty} \rho(s)ds \leq 2^{-\alpha K \lambda},
\end{equation}
where $\rho\in L^1(\R)$ is even symmetric on $\R$, monotonically decreases on $\R^+$, and
bounds $|\varphi|$ from above everywhere. This quantity can be interpreted as the margin that needs to be added to control the tail behavior of $\varphi$.
For this reason, we consider the larger `padded' interval $\widetilde{I} = [-a - T_0(\lambda), a + T_0(\lambda)]$, its dilation $\lambda \widetilde{I} = [-\lambda(a + T_0(\lambda)), \lambda(a + T_0(\lambda))]$, and the truncated approximation

\begin{equation}
\widetilde{f}_\lambda(t) =\frac{1}{\lambda} \sum\limits_{\Z\cap \lambda \widetilde{I}}
q_n^{\lambda} \varphi\left(t-\frac{n}{\lambda}\right).
\end{equation}

Restricting to $t \in I$, this function is close to any possible extension of the form $\widetilde{x}_{\lambda} = \frac{1}{\lambda} \sum\limits_{n \in \Z} q_n^{\lambda} \varphi\left(t-\frac{n}{\lambda}\right)$. Indeed, for $n\in\Z\setminus \lambda \widetilde{I}$ one has $\left |t - \frac{n}{\lambda}\right| > T_0(\lambda)$, so that
\begin{equation}
|\widetilde x_\lambda(t)-  \widetilde{f}_\lambda(t)| \leq \frac{1}{\lambda}\sum\limits_{\Z\setminus \lambda \widetilde{I}}
\left|\varphi\left(t-\frac{n}{\lambda}\right)\right|\leq 2 \int\limits_{T_0(\lambda)-\frac{1}{\lambda}}^{\infty} \rho(s)ds \leq 2^{-\alpha K \lambda+1};\label{eq:rhoint}
\end{equation}
Recall that $E^\mu_{K}(\lambda) = \sup_{x \in {\mathcal{B}_1(\R, \mu)}} \inf_{q_n^{\lambda} \in {\cal A}_K} \| x - \widetilde{x}_{\lambda} \|_{L^\infty(\mathbb{R})}  =  \leq C2^{-\alpha K \lambda}$ is assumed.  Then
\begin{align}
\|x  - \widetilde f_\lambda\|_{L^\infty(I)} &\leq  \|x-\widetilde x_\lambda\|_{L^\infty(I)}+ \|\widetilde x_\lambda-\widetilde f_\lambda\|_{L^\infty(I)}\\
& \leq C' 2^{-\alpha K \lambda}=:\varepsilon.
\end{align}
That is, for this choice of $\varepsilon$, the $\widetilde{f}_\lambda$'s form an $\varepsilon$-net for the space $\mathcal{B}_1(I, \mu)$.  It is clear that as $x$ varies in the set $\mathcal{B}_1 (\R,\mu)$, the resulting $\varepsilon$-net $F_{\lambda}$ has cardinality at most  $2^{K |\Z\cap\lambda \widetilde{I}|}$.

\subsection{An $\varepsilon$-net for the reduced space $\mathcal{B}^{\delta}_1(I, \mu)$}
By our main assumption there is a fixed constant $\alpha_0$ such that  $\alpha > \alpha_0 > 1 - K^{-1} (1 - h(\frac{1+\mu}{2}))$.  By continuity of $h$, we may fix $\delta > 0$ sufficiently small that $\alpha_0 \geq 1 - K^{-1}(1 - h(\frac{1}{2} + \frac{\mu - 5\delta}{2} ))$. For this choice of $\delta$, we will now estimate the size of the $\varepsilon$-net $F_{\lambda}^{\delta}$ arising in the same way as $F_\lambda$ when $x$ varies only over $\mathcal{B}^\delta_1 (\R,\mu)$. Note that $\delta$ may depend on $\mu$ but is independent of $\lambda$. Hence we can we assume without loss of generality that $\lambda$ is large enough to ensure $\varepsilon\leq \delta$. We note that for all $t$ one has $x(t)\geq \mu- \delta$, thus $\widetilde x_\lambda(t)\geq \mu- \delta-\varepsilon\geq \mu- 2\delta$, and, by (\ref{eq:rhoint}),  $\widetilde f_{\lambda}(t)\geq \mu-3\delta$ for $t \in I$.
Consequently, 
\begin{equation}
F_\lambda^\delta \subset F_\lambda \cap \big\{ f\in \mathcal{B}_1(\R, \mu) \hspace{2mm} | \hspace{2mm} \forall t\in I: f(t)\geq \mu- 3\delta \big\}.
\end{equation}
Let $G_{\lambda} = \big[ {\cal A}_K \big]^{\mathbb{Z} \cap \lambda \widetilde{I}}$, and consider the subset of this class given by
\begin{equation}
G_\lambda^\delta:=\left\{q_n^{\lambda} \in G_{\lambda}: \forall t\in I: \frac{1}{\lambda} \sum\limits_{\Z\cap \lambda \widetilde I}
q_n^{\lambda} \varphi\left(t-\frac{n}{\lambda}\right)\geq \mu-3\delta\right\}.
\end{equation}
 Now consider a random variable ${\bf Q}$ distributed according to a uniform probability measure on $G_\lambda$. We observe that
\begin{equation}\label{eq:rade}
| F_{\lambda}^\delta |\leq | G_{\lambda}^\delta |= \P({\bf Q} \in G_\lambda^\delta)  | G_\lambda |= \P({\bf Q} \in G_\lambda^\delta)  2^{K|\Z\cap\lambda \widetilde I |} \leq \P({\bf Q} \in G_\lambda^\delta)  2^{\lceil \lambda K (|I|+2 T_0(\lambda))\rceil}.
\end{equation}
We would now like to estimate $\P({\bf Q} \in G_\lambda^\delta)$:
\begin{enumerate}
\item Note that $\bf Q$ agrees in distribution with a sequence of identically distributed independent variables $Q_n$, $n\in \Z\cap\lambda \widetilde I$, which have support on $[-1,1]$ and expectation $\E\big( Q_n \big)= 0$. Consequently, one obtains
\begin{align}
\label{eq:bound}
\P({\bf Q} \in G_\lambda^\delta)\leq& \P\left(\forall j\in \Z \cap \lambda I:\frac{1}{\lambda} \sum\limits_{n\in \Z\cap \lambda \widetilde I}
Q_n \varphi\left(\frac{j-n}{\lambda}\right)\geq \mu-3\delta \right)\nonumber \\
\leq& \P\left(\frac{1}{\lambda}\sum_{j\in \Z \cap \lambda I } \sum\limits_{n\in\Z\cap \lambda \widetilde I}
Q_n \varphi\left(\frac{j-n}{\lambda}\right) \geq |\Z\cap \lambda I | (\mu-3\delta) \right) \nonumber \\
=&  \P\left(\sum\limits_{n\in \Z\cap \lambda \widetilde I} c_n Q_n \geq  |\Z\cap \lambda I | (\mu-3\delta) \right),
\end{align}
where $c_n = \frac{1}{\lambda}\sum_{j\in \Z \cap \lambda I }  \varphi\left(\frac{j-n}{\lambda}\right)$.
\item We would like to bound the coefficients $c_n$.  By assumption, we have $\varphi \in {\cal S}$.  Therefore, we may apply the Poisson Summation Formula, 
\begin{equation}
\frac{1}{\lambda}
\sum_{j\in \Z }  \varphi\left(\frac{j-n}{\lambda}\right) = \sum_{k \in \mathbb{Z}} \widehat{\varphi}(k \lambda)  = \widehat{\varphi}(0) + O( \lambda^{-1} ) = 1 + O( \lambda^{-1} ).
\end{equation}

From now on, we assume that $a > T_0(\lambda)$. This assumption makes sense as in the definition of the average Kolmogorov $\varepsilon$-entropy, one lets $|I|\rightarrow \infty$ for each fixed $\lambda$. Then the interval $\widehat{I} :=  [-(a - T_0(\lambda)), a - T_0(\lambda)]$ is non-empty; it satisfies  $\widehat{I} \subset I \subset \widetilde I$. 

Now for $n\in \lambda \widehat{I}$, we have
\begin{equation}
\left|\frac{1}{\lambda}\sum_{j\in \Z \cap \lambda I}  \varphi\left(\frac{j-n}{\lambda}\right)  - 1 \right| \leq 2\delta + O( \lambda^{-1} )
\end{equation}
as a consequence of (\ref{eq:T0}). Furthermore, for $n\notin \lambda \widehat{I}$, we use the crude estimate 
\begin{equation}
\left| \frac{1}{\lambda}\sum_{j\in \Z \cap \lambda I}  \varphi\left(\frac{j-n}{\lambda}\right) - 1 \right| \leq \|\rho\|_1 + 1+ \rho(0) =: D
\end{equation}
in conjunction with the bound
\begin{equation}
|\Z\cap(\lambda \widetilde I \setminus \lambda \widehat I)| \leq 4 \lambda T_0(\lambda)+4=: N (\lambda).
\end{equation}
\item We now apply these bounds for the coefficients $c_n$ in \eqref{eq:bound}.  As $\widehat{I} \subset I$, and $| \widetilde{I} | = O(\lambda)$, we obtain
\begin{align}
\P({\bf Q} \in G_\lambda^\delta)\leq &\P\left( \sum\limits_{\Z\cap \lambda \widetilde{I}} Q_n \geq  |\Z\cap \lambda I | (\mu-5\delta - O(\lambda^{-1}))  - D N(\lambda) \right) \\
\leq &\P\left(\sum\limits_{\Z\cap \lambda I} Q_n \geq  |\Z\cap \lambda I |(\mu-5\delta) - D' N(\lambda) \right).
\end{align}

Rescaling the random variables $Q_n$ to yield independent and identically distributed random variables supported on $[0,1]$ with expectation equal to $1/2$, we  may apply Propositions ~\ref{prop:max} and ~\ref{prop:1d} to bound the probability of such a large deviation from the mean:
\begin{align}
\log_2 \P({\bf Q} \in G_\lambda^\delta) \leq&-|\Z\cap \lambda I | h\Big(.5\big(1 + \mu-5\delta + C' N(\lambda) )\cdot | \Z\cap \lambda I |^{-1} \big) \Big) \\
\leq & -(\lambda |I|-1) \Big( h\Big(.5 \big( 1+ \mu-5\delta \big) \cdot \big(1+O(|I|^{-1}) \big) \Big) - 1 \Big).
\end{align}
\end{enumerate}

Combining our estimate for $ \P({\bf Q} \in G_\lambda^\delta)$ with (\ref{eq:rade}), we obtain that as $x$ varies in $\mathcal{B}_{1}^\delta (\R,\mu)$, the $f_\lambda$ vary on a set $F^\delta_\lambda$ of cardinality at most
\begin{equation}
\label{N}
N := 2^{ \lambda |I| \big(K- 1 +  h(\frac{1}{2}+\frac{\mu-5\delta}{2}) \big)\cdot ( 1+ O(|I|^{-1}) )}
\end{equation}

As a consequence, for each $\lambda > 1$, there are arbitrarily long intervals $I$
such that, for each $I$, there is an $\varepsilon$-net of $\mathcal{B}^\delta_{1} (I,\mu)$ with at most $N$ elements.  

\subsection{Towards a contradiction} We have found that the size of a $\varepsilon$-net of ${\mathcal B}_1^{\delta}(I,\mu)$ is bounded about by $N$, as given by equation \eqref{N}. Thus, we may bound the Kolmogorov $\varepsilon$-entropy $H_{\varepsilon}$ of the space ${\mathcal B}_1^{\delta}(I,\mu)$ (see  section $4.2$) by
\begin{equation}
 H_\varepsilon(|I|) \leq \log_2(N) = \lambda |I|\Big( K - 1 + h \Big( \frac{1}{2}+\frac{\mu-5\delta}{2} \Big)  \Big) \big( 1+ O(|I|^{-1}) \big).
\end{equation}
As $| I | \rightarrow \infty$, we may bound the average Kolmogorov $\varepsilon$-entropy $\bar{H}_{\varepsilon} = \lim_{| I | \rightarrow \infty} \frac{H_{\epsilon}(|I|)}{|I|}$ by $\bar{H}_{\varepsilon} \leq \lambda \Big(K - 1 +  h(\frac{1}{2}+\frac{\mu-5\delta}{2}) \Big)$.   Note that this also gives a bound on the average Kolmogorov $\varepsilon$-entropy for the larger  space ${\mathcal B}_1$.   Recalling that $\varepsilon = C' 2^{-\alpha K \lambda}$, or $\log_2\frac{1}{\varepsilon} =\alpha K \lambda - \log_2 C'$, and also recalling our hypothesis that $\alpha > \alpha_0 \geq 1 - K^{-1} + K^{-1} h( \frac{1+\mu-5\delta}{2} )$, we arrive at the chain of inequalities
\begin{equation}
\alpha -\frac{\log_2 C'}{K \lambda} \leq \left(\log_2\frac{1}{\varepsilon}\right) \frac{1- K^{-1} + K^{-1}h \big(\frac{1}{2}+\frac{\mu-5\delta}{2} \big)}{\bar{H}_{\varepsilon}} \leq  \left(\log_2\frac{1}{\varepsilon}\right) \frac{\alpha_0}{\bar{H}_{\varepsilon}}
\end{equation}
This establishes a contradiction to (\ref{eq:heps}), as $\lambda\rightarrow \infty$ and consequently $\varepsilon\rightarrow 0$. \qed

\paragraph{Remark} The assumption that the kernel $\varphi \in {\cal S}$ in our main theorem is stronger than necessary.  We used this assumption only to apply the Poisson Summation Formula in the proof of Theorem \ref{thm:main}; a weaker but more technical requirement for this formula to hold is that $| \varphi(t)|  + | \widehat{\varphi(t)} | \leq C( 1 + t )^{-(1 + \delta) }$.  In particular, our proof  also works  for twice continuously differentiable kernels $\varphi$ with compact support, a scenario resembling filters used in practice.




\section*{Acknowledgments}
The authors would like to thank Sinan G\"unt\"urk for interesting discussions on this topic as well as the Hausdorff Center for Mathematics, Bonn, and the Summer School on ``Theoretical Foundations and Numerical Methods for Sparse Recovery'' at the RICAM, Linz, where large parts of the project were completed.  They gratefully acknowledge the support of a National Science Foundation Postdoctoral Research Fellowship (Ward) and the Charles M. Newman Fellowship at the Courant Institute (Krahmer).

\bibliography{model1-num-names}

\begin{thebibliography}{10}

\bibitem{DC02}
A.~Calderbank and I.~Daubechies.
\newblock The pros and cons of democracy.
\newblock {\em IEEE Transactions on Information Theory}, 48:1721--1725, 2002.

\bibitem{DGK10}
P.~Deift, S.~G\"unt\"urk, and F.~Krahmer.
\newblock An optimal family of exponentially accurate one-bit sigma-delta
  quantization schemes.
\newblock submitted.

\bibitem{SQG}
M.~Derpich, E.~Silva, D.~Quevado, and G.~Goodwin.
\newblock On optimal perfect reconstruction feedback quantizers.
\newblock {\em IEEE Transactions on Signal Processing}, 56:3871--3890, 2008.

\bibitem{Ellis}
R.~Ellis.
\newblock {\em Entropy, Large Deviations and Statistical Mechanics}.
\newblock Springer Publication, 2005.

\bibitem{wireless}
I.~Galton.
\newblock Delta-sigma data conversion in wireless transceivers.
\newblock {\em Microwave Theory and Techniques, IEEE Transactions on},
  50(1):302 --315, Jan 2002.

\bibitem{AG}
L.~Gordon and R.~Arratia.
\newblock Tutorial on large deviations for the binomial distribution.
\newblock {\em Bulletin of {M}athematical {B}iology}, 51(1), 1989.

\bibitem{S}
C.~G\"unt\"urk.
\newblock One-bit sigma-delta quantization with exponential accuracy.
\newblock {\em Communications on Pure and Applied Mathematics},
  (11):1608--1630, 2003.

\bibitem{KT}
A.~Kolmogorov and M.~Tihomirov.
\newblock $\epsilon$-entropy and $\epsilon$-capacity of sets in function
  spaces.
\newblock {\em Uspehi Mat Nauk}, (2):3--86, 1959.

\bibitem{LP}
C.~Leon and F.~Perron.
\newblock Extremal properties of sums of {B}ernoulli random variables.
\newblock {\em Statistics and Probability Letters}, 2003.

\bibitem{SD}
S.~Norsworthy, R.~Schreier, and G.~Temes.
\newblock {\em Delta-Sigma Data Converters: Theory, Design and Simulation}.
\newblock New York: IEEE Press, 1997.

\end{thebibliography}
\bibliographystyle{abbrv}

\end{document}